\documentclass[prb, showpacs, preprint, preprintnumbers]{revtex4}
\usepackage{graphicx}

\begin{document}

\section{}

\title{Attractive interaction of  Indium with point defects and silicon in GeSi }
\author{ R. Govindaraj  and R. Sielemann}
\affiliation{Hahn-Meitner Institute,Glienicker Strasse,100, Berlin, Germany}
\date{\today}
\begin {abstract}
In electron irradiated Ge$_{0.98}$Si$_{0.02}$, In-vacancy and In-interstitial complexes were 
observed in analogy with defects  in pure Ge. Isochronal annealing measurements 
reveal that the temperature of dissociation of In-defect complexes deviate from 
pure Ge subjected to identical e-irradiation, which is explained on the basis of strain induced by undersized silicon 
atoms affecting the binding energy of In-V and In-I complexes. Besides the pairing with 
intrinsic defects the interaction of In with Si atoms is observed resulting in several 
different configurations. Complementary experiments performed in Ge$_{0.98}$Si$_{0.02}$
and Ge$_{0.94}$Si$_{0.06}$ elucidate the attractive  interaction between In and Si.

\end{abstract}
\maketitle 

Germanium-silicon (Ge$_x$Si$_{1-x}$) alloy, 
is an exciting material  for 
band gap engineering and the integration of fast analog 
circuits \cite{1,2,3}, due to tunable 
lattice constant and band gap energy with  x. Interactions of indium (an important
acceptor impurity in Ge/ Si system) with point defects 
have been carried out  in detail in Ge, as a function of 
electronic chemical potential by Perturbed angular correlation (PAC) spectroscopy  \cite{4,5}. 
PAC method has been applied  to defect studies in elemental and compound semiconductors \cite{6}.
Of special importance for the PAC study of semiconductors is the 
need for only a dilute concentration of probe atoms (typically 10$^{13}$ cm$^{-3}$
to 10$^{14}$ cm$^{-3}$) and the possibility of  studying the sample at different temperatures. 
The present PAC experiment  illustrates the interaction between the electron 
irradiation induced defects and indium atoms  in Ge$_{0.98}$Si$_{0.02}$  ,  thus   elucidating  the 
role of  2 at\% silicon in the formation and evolution of point defects.\\

Ge$_{0.98}$Si$_{0.02}$  crystals (n-type and n$_e$= 10$^{15}$ cm$^{-3}$) were  grown  along $<110>$  by Czochralski technique
as reported  \cite{7}. 
$^{111}$In concentration $\le$ 5x10$^{13}$ cm$^{-3}$ of probe atoms
are recoil implanted uniformly to a depth of 4 $\mu$m as reported earlier \cite{4,5}.
The coincidence spectra between 171 and 245 keV $\gamma$ rays of $^{111}$Cd were measured 
using a four BaF$_2$ detector PAC spectrometer with a time resolution of 660 ps \cite{5}.  
The measured  perturbation function 
R(t) \cite{7} were analysed for static quadrupole interactions as
R(t)= A$_2$ ( f$_o$ exp(-$\Delta_0 t) + \sum_{i=1}^n f_i G_i(t)$), with the
perturbation factor G(t) is given as  $\sum_{m=0}^3$ S$_{2m}$($\eta$) cos(g$_n$ ($\eta$) $\omega_m$t) exp(-g$_n$($\eta$)$\delta_m$$\omega_m$t)
where n is determined by the number of frequency components occuring in the R(t) spectrum.
The frequencies $\nu_Q$ are related to $\omega_m$ by $\omega_m$ = (m 3 $\pi$ / 10 ) $\nu_Q$,
when the interaction is axially symmetric. 
The parameters evaluvated are  ${viz}$., quadrupole frequency $\nu_Q$ (=eQV$_{zz}$/h),   
width of Lorentzian distribution of  quadrupole frequencies  $\delta$ 
as experienced by the fraction f of probe atoms \cite{5}. A$_2$ is the effective anisotropy of the
$\gamma-\gamma$ cascade and $\Delta_o$ denotes the width of the Lorentzian distribution of
quadrupole frequencies with a mean at 0 MHz. While $\delta_m$ = $\Delta \nu_{Qm}/ \nu_{Qm}$,
where $\Delta \nu_{Qm}$ is the spread of interaction strength. 
PAC measurements were carried out in $^{111}$In recoil implanted Ge$_{0.98}$Si$_{0.02}$.
Subsequent to annealing at 823 K, the sample was irradiated with electron of energy 
2 MeV  to a dose of  10$^{15}$ e/cm$^2$ at 77 K to produce  frenkal pairs in a controlled 
manner and isochronal annealing measurements have been carried out. Measurements have also been 
carried out in $^{111}$In recoil implanted Ge$_{0.94}$Si$_{0.06}$ to further elucidate the role 
of Si.\\
The quadrupole parameters in $^{111}$In recoil implanted sample at
room temperature are listed in Table-I. Based on the
reported results \cite{4,5,8,9}, f$_1$ and f$_2$ are identified to be due to 
In-vacancy (52 MHz,$\eta$= 0) and In-interstitial (415 MHz,$\eta$ = 0) 
complexes respectively. Computationally these are deduced to be
split vacancy with Cd (on the bond centre) and Cd-self-interstitial complex
with split configuration respectively \cite{10}.
A sharp slope in the R(t) spectrum 
(Cf Fig.1a) around t$\approx$ 0 ns  is    
mostly contributed by $^{111}$Sn recoil implanted  defects located at next nearest neighboring 
environment of the probe atoms. PAC 
spectrum in the recoil implanted Ge$_{0.98}$Si$_{0.02}$ (Fig.1a)  is 
distinctly different from that of Ge \cite{11,13}.
In $^{111}$In implanted n-Ge (with charge carrier concentration 
n$_e$ $\approx$ 10$^{15}$ e$^-$/cm$^{3}$) it is observed that interstitials are
predominantly present compared to vacancies. In the present sample (n-Ge$_{0.98}$Si$_{0.02}$ of
similar n$_e$) we observe that both vacancies and interstitials are predominantly present (Cf. Table-I), thus
bringing out the significant role played by silicon in stabilising vacancies at the
cost of interstitials, the details of which are discussed subsequently.\\

The sample is annealed at 888 K  for 15 minutes (Cf. Fig 1b) to  restore all the probe atoms at 
defect free substitutional sites \cite{4}. 
The fractions f$_3$ 
and f$_4$ are plausibly interpreted as  In-Si$_2$ complex and In-Si clusters, the discussion on these assignemnt
is presented in the later part of this work. 

Results of PAC measurements in the e-irradiated 
sample are compiled in Table-I. Corresponding PAC spectra are shown in Fig 1. 
In Ge$_{0.98}$Si$_{0.02}$ , the fractions f$_1$ (In-V$_0$) and f$_2$ (interstitial-In) disappear
following annealing treatments at 450 and 350 K respectively (Cf. Fig.2) , whereas
in pure Ge subjected to electron irradiation
it has been observed that  In-I$_0$,  In-V$_0$  complexes dissociate around 380 and 400 K 
respectively \cite{5}.  These results imply an accelerated  recovery of  In-I$_0$ complex and 
a  retarded dissociation  of In-V$_0$ complex in the case of Ge$_{0.98}$Si$_{0.02}$ compared with 
pure Ge subjected to identical electron irradiation.  The difference in recovery stages 
can be understood due to the strain  developed at undersized silicon atoms (r$_{Ge}$ = 1.22 A$^o$ and r$_{Si}$ = 1.17 A$^o$).  The strain developed 
at Si might have an influence over a few neighbors affecting the displacement fields of 
interstitial and vacancies thus resulting in a  change of binding energy of In-V$_0$ and 
In-I$_0$ complexes. Silicon being undersized than Ge  would have more binding to interstitial 
complex contributing for the lower binding energy of In-I$_0$ complex in Ge$_{0.98}$Si$_{0.02}$ 
than in pure Ge,  while the same effect contributes for an increased binding energy 
of In-V$_0$ complex.  
\begingroup
\squeezetable
\begin{table*}
\caption{Results of TDPAC measurements in Ge$_{0.98}$Si$_{0.02}$ and Ge$_{0.94}$Si$_{0.06}$ }
\begin{ruledtabular}
\begin{tabular*}{120mm}{|c|c|c|c|c|c|}
Sample treatment&index& $\nu_{Qi}$&~$\eta_i$~& f$_i$       &identification \\
                &     & (MHz)     &          &             &               \\
                &     &           &          &             &               \\
\hline
$^{111}$Sn recoil implanted & 0 & 0       & 0 & 0.83$\pm$0.01    &substitutional probe atoms \\
Ge$_{0.98}$Si$_{0.02}$      & 1 & 52$\pm$1& 0  &0.10$\pm$0.02    &                 In-V$_O$\\
                            & 2 & 415$\pm$2& 0 &0.07$\pm$0.01    &                 In-I$_O$\\
\hline
$^{111}$Sn recoil implanted & 0 & 0       & 0                     & 0.64$\pm$0.03         & substitutional probe atoms \\
and annealed (at 880 K) Ge$_{0.98}$Si$_{0.02}$ & 3 &11$\pm$2& 0.3 &  0.31$\pm$0.03  & In-Si$_2$ $/$ In-Si complex\\
                            & 4 &34$\pm$3& 0       & 0.05$\pm$0.01                   & In-Si clusters\\
\hline
electron irradiated         & 0 & 0       & 0          &  0.59$\pm$0.02      & substitutional probe atoms \\
Ge$_{0.98}$Si$_{0.02}$      & 1 & 52$\pm$1& 0          &  0.12$\pm$0.01              &     In-V$_O$\\
                            & 2 &415$\pm$2& 0          &  0.09$\pm$0.01              &       In-I$_O$ \\
                            & 3 &15$\pm$3 & 0.35       &  0.2$\pm$0.03              &       In-Si$_2$ $/$ In-Si complex\\
\hline
electron irradiated         & 0 & 0       & 0          & 0.64$\pm$0.03      & substitutional probe atoms \\
Ge$_{0.98}$Si$_{0.02}$      & 1 & 52$\pm$1& 0          &  0.11$\pm$0.01       &     In-V$_O$\\
and annealed at 338  K      & 2 &415$\pm$2& 0          &  0.05$\pm$0.01       &       In-I$_O$ \\
                            & 3 &15$\pm$3 & 0.35       &  0.2$\pm$0.03      &       In-Si$_2$ $/$ In-Si complex\\
\hline
electron irradiated         & 0 & 0       & 0          &  0.74$\pm$0.03      & substitutional probe atoms \\
Ge$_{0.98}$Si$_{0.02}$       
and annealed  at 880 K      & 3 & 11$\pm$2 &0.32       &  0.2$\pm$0.02      &    In-Si$_2$ $/$ In-Si complex\\
                            & 4 &96$\pm$2 & 0          &  0.06$\pm$0.01     &     In-Si clusters\\
\hline
$^{111}$Sn recoil implanted & 0 & 0       & 0          & 0.48$\pm$0.04     &   substituional probe atoms \\
and annealed (at 880 K)  Ge$_{0.94}$Si$_{0.06}$ & 3 & 10$\pm$2& 0.4     & 0.43$\pm$0.02   & In-Si$_2$ $/$ In-Si complex\\
                            & 4 &62$\pm$3& 0 &  0.09$\pm$0.02                             & In-Si clusters\\
\hline
e$^-$ irradiated and annealed (at 880 K) Ge &  & 0       & 0   &  1    &  substituional probe atoms 
\end{tabular*}

\end{ruledtabular}
\end{table*}
\endgroup
The fractions f$_3$ and f$_4$  are absent in the starting sample (in which probe atoms were 
recoil implanted)  but only occurs  following the annealing 
at 850 K.  This observation rules out their formation due to  athermal migration of defects. 
Also these fractions were not observed in pure Ge sample subjected to identical electron irradiation 
and annealed at 850 K \cite{5}, thus implying the association of Si atoms with these
complexes. A high value of f$_3$ and non zero  $\eta_3$ (Cf. Table-I)
imply a simple nature of the defect but axially asymmetric. Hence
the complex  cannot be In-Si, as In-X (where X= P,Sb etc.,) pairs are mostly axially symmetric \cite{14}.  
Therefore f$_3$ is  plausibly  interpreted to be In-Si$_n$ complex with n $\ge$ 2. 
This is analogous to the formation of Ge dimers aided by 
divacancies in electron irradiated SiGe systems \cite{16}.    
f$_4$ with axially symmetric configuration  is interpreted to be due to In-Si 
clusters.  The formation of In-Si complexes /clusters  leading to the occurrence of f$_3$  
and f$_4$ could be contributed by irradiation  induced defects with an 
important role being played by strain at silicon atoms.  Since  f$_3$ and f$_4$ 
were observed in the probe recoil implanted sample following annealing at 850 K, 
we would explore  the role of  the diffusion of silicon atoms in the
formation of In-Si complexes.  Assuming the diffusion coefficient of silicon in 
germanium to be around 10$^{-17}$ cm$^2$/sec around 823 K \cite{16} , 
the diffusion length can be computed to be around  40 A$^o$ for the (experimental) 
annealing time of 3600 seconds.   In the sphere of 40 A$^{o}$ radius about 200 silicon 
atoms are present in Ge$_{0.98}$Si$_{0.02}$, implying  an appreciable 
probability for the formation of In-Si complexes. In addition 
the strain developed due to undersized silicon could enhance diffusion \cite{16} leading to a higher
probability for the formation of such complexes.  
The occurrence of  In-Si complexes 
and In-Si clusters indicate attractive interaction between In and Si atoms. 
Now we will discuss the effect of Si on the electronic
properties of the samples (beyond the defect recovery stages)
in terms of the experimental quadrupole parameters. R(t) spectrum (Cf. Fig 4a)  
corresponding to electron irradiated and well annealed (at 880 K) pure n-Ge
at 10 K. In the absence of any 
quadrupole frequency component,the  Dampening parameter ($\Delta_0$) associated with f$_0$ as
deduced in this case is 2.3 MHz similar to the temperature dependence of hyperfine 
interactions of Cd in Ge \cite{11}. Details of various electronic effects contibuting
for the occurrence of $\Delta_0$ is discussed elsewhere \cite{11}. Analysis of the PAC spectra in well annealed  Ge$_{0.98}$Si$_{0.02}$ and Ge$_{0.94}$Si$_{0.06}$ 
samples (Cf. Fig 4b and 4c) at 300 K, show that the value of $\Delta_0$ are
around 3.2 and 3.7 MHz respectively. The observed increase in $\Delta_0$ with increasing silicon concentration  
with the PAC measurements carried out at 300 K is  understood based 
on electronic effects preferably due to higher JT distortion. Summarizing, in electron irradiated Ge$_{0.98}$Si$_{0.02}$ 
there is an accelerated (retarded) recovery of  In-I$_0$  (In-V$_0$ ) complexes. The 
occurrence of In-Si complexes and In-Si clusters 
in Ge$_{0.98}$Si$_{0.02}$ and 
Ge$_{0.94}$Si$_{0.06}$ indicates a strong attractive interaction between 
In and Si. \\
\vskip 10pt
Results reported here are based on preliminary measurements carried out on GeSi.
Detailed PAC measurements in GeSi with slightly different silicon composition 
and subsequent to a larger number of annealing steps have to be carried out for a
complete and more detailed understanding of the aspects related to In-defect interactions as 
influenced by silicon in GeSi.
\vskip 12pt
We thank Peter symkoviak for all the help and Drs. Schroder $et~al$ \cite{7} for providing us single crystalline samples.
This work was carried out by both the authors at HMI while RG was on leave of absence 
from IGCAR, Kalpakkam. 
\newpage
{\bf Figure captions}\\
\vskip 12pt
Fig 1. {TDPAC spectra in Ge$_{0.98}$Si$_{0.02}$ for the following cases $viz$
(a) as $^{111}$Sn recoil implanted (b) annealed at 880 K (c) e$^-$ irradiated
(d) annealed at 338 K (e) 425 K and (f) 923 K}
\vskip 4pt
Fig 2. {Variation of f$_1$ (In-V$_0$;52 MHz) and f$_2$ (In-I$_0$;415 MHz)
with annealing temperature}
\vskip 4pt
Fig 3. {Variation of hyperfine parameters of In-Si complexes with
annealing temperature }
\vskip 4pt
Fig 4. {TDPAC spectra corresponding to annealed (at 880 K) (a) Ge (b) Ge$_{0.98}$Si$_{0.02}$
and (c) Ge$_{0.94}$Si$_{0.06}$}


\vskip 24pt
\newpage
\begin{references}
\bibitem{1}
{B.S. Meyerson, {\it Scientific American} {\bf 9} , 42 (1994).}
\bibitem{2}
{J. weber, {it Phys  Rev} {\bf B40}, 5683 (1989).}
\bibitem{3}
{Temkin ${et~al}$,  {\it Appl. Phys. Lett.} {\bf 52}, 1089 (1988)}
\bibitem{4}
{M. Brussler, H. Metzner and R. Sielemann, {\it Materials Science Forum} {\bf 38}, 1205  (1989)}
\bibitem{5}
{H. Haesslein, R. Sielemann and C. Zistl, {\it  Phys. Rev. Lett}, {\bf 80}, 2626 (1998)}
\bibitem{6}
{Th. Wichert, N. Achtziger, H. Metzner and R. Sielemann, in 
{\it Hyperfine interactions of Defects in Semiconductors}, 
edited by G. Langouche (Elsevier, Amsterdam, 1992), p79.}
\bibitem{7}
{N. V. Abrosimov {\it et~al}, {\it J. Crys. Growth} {\bf 174}, 182 (1997)}
\bibitem{8}
{R. Sielemann,  {\it Nucl. Instr. Meth.} {\bf B146} 329 (1998)}
\bibitem{9}
{R. Sielemann {\it et~al}, {\it Physica} {\bf B273-274} 565 (1999)}
\bibitem{10}
{H. Hohler {\it et al}, {\it Phys. Rev.} {\bf B70} 155313 (2004)}
\bibitem{11}
{A. F. Pasquevich and R. Vianden, {\it Phys. Rev.} {\bf B37}, 10858 (1988)}; Also see
{A. F. Pasquevich and R. Vianden, {\it Phys. Rev.} {\bf B41}, 10956 (1990)}
\bibitem{12}
{A. da Siliva {\it et~al}, {\it Phys. Rev.} {\bf B62} 9903 (2000)}
\bibitem{13}
{A. Fazzio {\it et~al}, {\it Phys. Rev.} {\bf B61} R2401 (2000)}
\bibitem{14}
{N. Achtziger, W. Witthuhn, {\it Phys. Rev.} {\bf B47} 6990 (1993)}
\bibitem{15} 
{J. W. Corbett and J. C. Bourgoin, in {\it Point Defects in Solids}, edited by J. H. Crawford and L. M. Slifkin (Plenum, Newyork, 1975) Vol 2}
\bibitem{16}
{S. M. Prokes, O. J. Glembocki and D. J. Godbey, {\it Appl. Phys. Lett.} {\bf 60} 1087 (1992)}
\end{references}
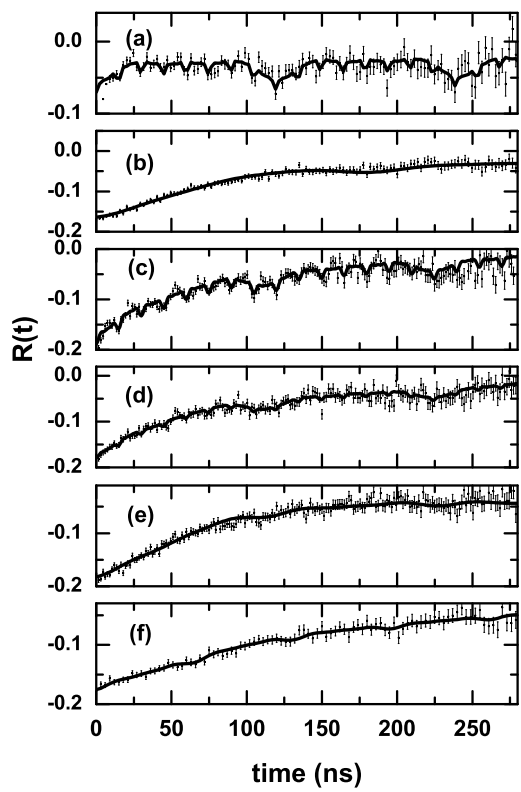
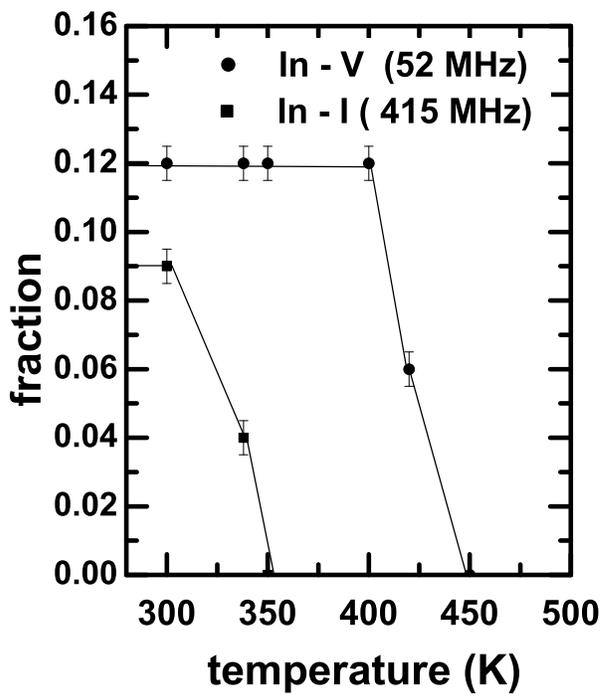
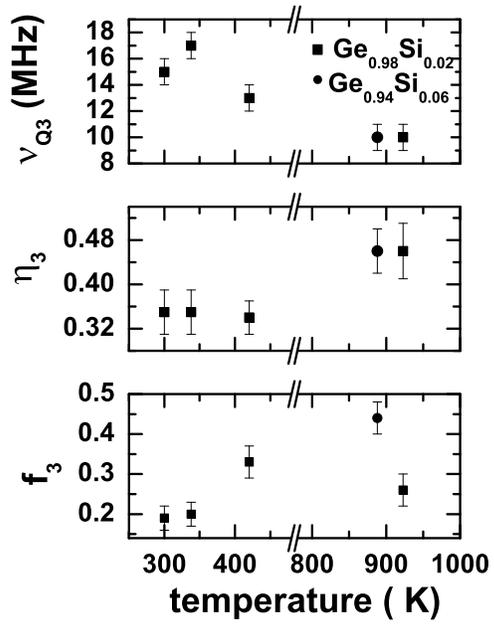
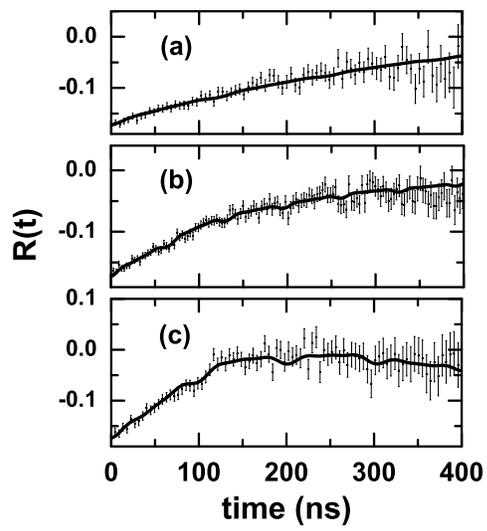
\end{document}